\documentclass[11pt]{article}
\usepackage{moriond,epsfig}

\def\be{\begin{equation}}
\def\ee{\end{equation}}
\def\bea{\begin{eqnarray}}
\def\eea{\end{eqnarray}}

\begin{document}
\vspace*{4cm}
\title{Redshift of Galaxy Clusters from the Sunyaev-Zel'dovich effect.}

\author{ J.M  Diego$^1$, J. Mohr $^2$, J. Silk$^1$, and G. Bryan$^1$.}

\address{$^1$Astrophysics Dept. University of Oxford. Keble Road, Oxford OX1 3RH,  UK\\ 
   $^2$Dept. of Astronomy and Physics, Univ. of Illinois at Urbana-Champaign, Urbana, IL 61801.}

\maketitle\abstracts{
We develop a new method for estimating the redshift of galaxy clusters 
through resolved images of the Sunyaev-Zel'dovich effect (SZE). Our method is based 
on morphological observables which can be measured by actual and future SZE experiments.   
The method is tested using a set of high resolution hydrodynamical simulations 
of galaxy clusters at different redshifts. The method combines the observables in a 
principal component analysis. We show how this can give an estimate of the 
redshift of the galaxy clusters. \\ 
Although the uncertainty in the redshift estimation is large, the method should be useful 
for future SZE surveys where hundreds of clusters are expected to be detected. 
A first preselection of the high redshift candidates could be done using our proposed morphological 
redshift estimator.}

\section{Introduction}
The advent of new experiments dedicated to the observation of the Sunyaev-Zel'dovich 
effect (Sunyaev \& Zel'dovich, 1972) (SZE hereafter), 
demands the development of new techniques to best analyze 
these new and exciting data. \\
Through the SZE it is possible to trace the hot plasma in the galaxy clusters which 
distorts the spectrum of the cosmic background radiation. 
This distortion is redshift independent and it is proportional to the temperature of 
the plasma and its electron density ($n_e$). This quality ($z$-independent, and $\propto n_e$) 
makes the SZE effect an ideal way to explore the high redshift population of galaxy clusters. 
However, the fact that the distortion induced by the cluster in the CMB is independent of the redshift 
of the cluster, makes the determination of the cluster redshift from SZE observations 
a challenging task. \\ 
Redshifts can be easily measured for relatively nearby clusters but for distant clusters one should use 
other approaches, for instance photometric redshifts. 
However, photometric redshifts for clusters above redshift $\approx$ 0.5 require large telescopes. 
Ongoing and future SZE experiments will detect hundreds or thousands of clusters. 
The redshift estimation of all these clusters using 10-m class telescopes is unfeasible. 
An optimal solution could be to combine small, and medium-sized telescopes to determine 
the redshifts of the low and intermediate $z$ clusters respectively. 
Then, we could leave the estimation of the redshifts of the most distant clusters to the large telescopes. 
However, to select the low, intermediate and high z clusters we need an estimate of their redshift. 
The motivation of this work is to show how it is possible to make this preselection of the 
low, intermediate and high redshift clusters using SZE data alone. Then, this preselection 
can be used to determine more accurately the redshifts of the clusters combining 
small (for the low z clusters) medium-sized (for the intermediate z clusters) 
and large (for the high z clusters) telescopes. \\
Our method is based only on observed quantities of the SZE. These quantities are associated with 
the observed shapes of the 2D surface brightness profile of the clusters which does have some 
dependence on the redshift. The observed size of a particular cluster, for instance, will 
decrease with increasing redshift. So, given an observed size we could, in principle, 
get a probability for the cluster to be at a given redshift. 
However, the size of the cluster will also depend on its total mass. 
This means that two clusters with different redshifts and masses could have the same apparent 
size provided the most distant cluster had a larger mass that compensates the decrease in the 
apparent size due to the increase in redshift. There are, therefore, degeneracies between 
the redshift of the cluster and its mass. The question now is, can we break 
this degeneracy by including more information in our analysis ? A resolved SZE image of a cluster 
provides information, not only about its size, but also about the shape of the entire 
profile. The total observed flux of the cluster, for instance, will depend on the total mass 
of the cluster (and its redshift and temperature). The central SZE decrement will depend 
on the core radius and electron central density but not on the redshift. 
By adding these and other additional observables, it should be possible to break the degeneracies between 
the mass and redshift. \\
Our method will have, however, one limitation: it works with resolved SZE images. 
Therefore it should be useful for sub-arcmin experiments but not for experiments like Planck 
where the best resolution will be 5 arcmin. \\
\section{Morphological redshifts}

The idea behind morphological redshift estimation is that by combining many observables 
taken from the 2D SZE cluster profile it is possible to {\it divide} the clusters in 
different groups, each one for a different redshift interval.  \\
The first two obvious observables are the total flux and size of the cluster. However, in real 
experiments it will be difficult to measure these two quantities due to the instrumental noise 
and the limited sensitivity of the detectors. In order to get the total flux and size of the 
cluster, one should extrapolate  the observed 2D profile below the noise level. 
The easiest approach to solve this problem is just to consider the isophotal size and 
isophotal flux instead of the total flux and total size of the cluster. 
Isophotal fluxes and sizes are direct observed quantities. 
Like the total flux and total size, their isophotal equivalents also show strong 
scaling relations in galaxy clusters (see Mohr \& Evrard 1997, Mohr et al. 2000, Verde et al. 2000 
for a relation between isophotal size and temperature).\\
If we only use the isophotal flux and isophotal size and we take into account that 
the scaling relations in galaxy clusters have an intrinsic scatter, we find that the redshift 
estimation is quite poor. In order to improve the result, more observables should be combined. \\
In this work we will include 5 additional observables.
The central amplitude, second derivative in the center, and 3 mexican hat wavelet coefficients 
at the center. \\
The central amplitude does not depend on the redshift but depends on the total 
projected mass along the line of sight. This will make this quantity very useful to break the degeneracy 
between the mass and the redshift when combined with other observables which do depend on the redshift. 
One such observable is the second derivative at the center. It gives an idea of how cuspy the 
cluster is. If we move a cluster of fixed mass back in redshift, we will see that the central 
amplitude does not change 
with redshift but the cluster becomes more and more cuspy. Related to the second derivative is the mexican 
hat wavelet. By changing the scale of this wavelet we can trace the 2D SZE profile at different radii. 
Although in this work we will present only the preliminary results obtained with these 7 observables, 
in a subsequent work (Diego et al.) we will show how it is possible to improve the result by including 
more observables. \\
In the next section we will show how to combine all these observables and use that combination to get 
an estimate of the redshift. \\

\subsection{Principal Component Analysis}
Principal component analysis (PCA hereafter) has been widely used in the last years as a 
powerful classifier of data sets. For our particular case, 
PCA has several desirable advantages which can be briefly summarized as follows. \\

\noindent
$\bullet$ PCA produces an optimal linear combination of the observables. It is optimal in the 
          sense that it maximizes the variance of the linear combination (or {\it projection}).\\

\noindent
$\bullet$ There is no limit in the number of observables. One can include as many observables 
          as wanted. \\

\noindent
$\bullet$ PCA is a {\it non-parametric method}. This is a key point since no assumptions about the 
          cluster scaling relations need to be done. If there are intrinsic scaling relations between 
          some of the observables, PCA will find them. \\

If we write our data set in matrix notation, $X_{Nm}$ ($N$ clusters observed and $m$ observables per 
cluster), then the principal components are given by the projections along the directions of the eigenvectors 
of the matrix $S= X^t X$. \\
The matrix $S$ has as many eigenvectors as observables (7 in our case). One quality of these eigenvectors 
is that they are independent one from each other. So, when projecting the data set, $X$, along the directions 
given by the eigenvectors we have a representation of the original data set in an orthogonal system. Another 
quality of these new representation of the data is that the projection along the direction of the eigenvector 
with the highest associated eigenvalue retains the highest percentage of information of the original 
data set. The eigenvector associated with the second highest eigenvalue retains the second largest percentage of 
information and so on. Therefore we can reduce the dimensionality of the problem by considering only the $p$ 
first directions (eigenvectors) such that the percentage of information retained by these $p$ directions 
is above certain threshold (typically the percentage should be $> 95 \%$). 
\section{Results}
\begin{figure}[h]
\begin{center}
\psfig{figure=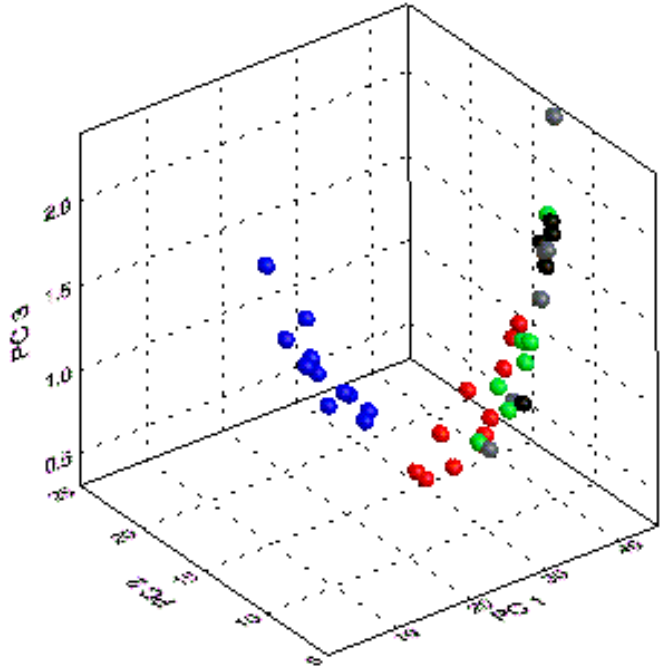,height=4.0in}
\end{center}
\caption{First three principal components. Blue points are clusters at redshift $\approx 0$, 
yellow at $z = 0.5$, grey at $z = 1$ and black at $z = 2.3$.}
\label{fig_1}
\end{figure}

We have applied the previous method to hydrodynamical N-body simulations of the SZE. 
We have filtered the images with a Gaussian of FWHM = 15 arcsec simulating the effect of an 
antenna and we have set a threshold in the maps which simulates the sensitivity 
of our instrument. \\
After computing the principal components we keep only the first three principal components 
since they will retain $\approx 95 \%$ of the variance of the original data set. 
In Figure \ref{fig_1} we show the result. As can be seen, different redshifts are grouped in 
different regions in this 3D space. This suggests that it will be possible to discriminate 
between low, intermediate and high redshift clusters. 
Our method should be useful for future sub-arcmin SZE surveys where hundreds of clusters should be 
detected and a preselection of high-intermediate-low redshift clusters should be very useful 
to optimize the optical follow up. 

In a subsequent work (Diego et al.) we will apply the method to better high resolution N-body simulations 
and we will include more observables in the analysis. 
Finally, we will recover the redshifts of our simulated clusters using a {\it Bayesian} approach 
and we will give the error of the recovered redshifts as a function of $z$.

\section*{Acknowledgments}
This research has been supported by a Marie Curie Fellowship 
of the European Community programme {\it Improving the Human Research 
Potential and Socio-Economic knowledge} under 
contract number HPMF-CT-2000-00967.

\section*{References}


\begin{thebibliography}{99}

\bibitem{Diego} Diego J.M., Mohr J.J, Silk J., Bryan G., in preparation.

\bibitem{Mohr} Mohr J.J., Evrard A., 1997, ApJ, 491, 38.

\bibitem{Mohr} Mohr J.J., Reese E.D., Ellingson E., Lewis A.D., Evrard A., 2000, ApJ, 544, 109.

\bibitem{Sunyaev} Sunyaev R.A., Zel'dovich Ya.B., 1972, A\&A, 20, 189.

\bibitem{Verde} Verde L., Kamionkowski M., Mohr J.J., Benson A.J., 2001,MNRAS, 321, L7.


\end{thebibliography}
\end{document}